\documentclass{eas}
\usepackage{graphicx}
\begin{document}

\title{Wandering globular clusters: the first dwarf galaxies in the universe?} 
\author{Myung Gyoon Lee$^1$, Sungsoon Lim}
\address{Department of Physics and Astronomy, Seoul National Unviersity, Seoul, Korea; \email{mglee@astro.snu.ac.kr}}
\author{Hong Soo Park}\address{Korea Astronomy and Space Science Institute, Korea}
\author{Ho Seong Hwang}\address{CEA/Saclay, France}
\author{Narae Hwang}\address{National Astronomical Observatory of Japan, Japan}
\begin{abstract}
In the last decade we witness an advent of new types of dwarf stellar systems including ultra-compact dwarfs, ultra-faint dwarf spheroidals, and exotic globular clusters, breaking the old simple paradigm for dwarf galaxies and globular clusters. 
These objects become more intriguing, and understanding of these new findings becomes more challenging.
Recently we discovered a new type of large scale structure in the Virgo cluster
of galaxies: it is composed of globular clusters. Globular clusters in Virgo are found wandering between galaxies (intracluster globular clusters) as well as in galaxies. These intracluster globular clusters fill a significant fraction in the area of the Virgo cluster and they are dominated by blue globular clusters.
These intracluster globular clusters may be closely related with the first dwarf galaxies in the universe. 
\end{abstract}
\maketitle

\section{Introduction}

In the framework of hierarchical structure formation based on $\Lambda$CDM models, dark matter and baryonic matter are located along the large scale filamentary structures, and galaxies and galaxy clusters are formed in these structures via various types of merging/accretion of smaller scale objects.  
This paradigm has been successful in explaining several aspects of modern observations on large scales.
  
However, this paradigm has been challenged by small-scale structure problems. 
One of them is the missing satellite problem that the number of known satellite dwarf galaxies in the Local Group is significantly smaller than predicted theoretically. To understand this problem it is important to make observational efforts to check whether there are some satellite dwarf galaxies waiting to be discovered in the Local Group as well as to make theoretical approach.

With the recent advent of very wide surveys, new types of dwarf galaxies are being discovered in the local universe, such as ultra-compact dwarf (UCD) galaxies, ultra-faint dwarf spheroidal galaxies, and dwarf to globular cluster objects (DTGOs). 
In addition, new types of star clusters also have been  found: globular clusters that have multiple stellar populations,
extended star clusters, faint fuzzy clusters, and isolated globular clusters located very far from the galaxies. 
Globular cluseters once believed to have a simple history are
turning out to have a complex history, and even some of them may be remnants of dwarf galaxies. 

Thus the universe of dwarfs is becoming more abundant and more diverse. Currently the nature of these objects is intriguing, and understanding of these intriguing objects becomes more challenging.
We introduce another kind to this universe of dwarfs:
wandering globular clusters.

\section{Wandering globular clusters}

In 1956 van den Bergh(1956) suggested from his finding that Palomar 4, globular cluster in the halo of our Galaxy, is at a distance of 145 kpc that it may be an
intergalactic tramp.
Recently these isolated star clusters are also found in the remote area in NGC 6822 (Hwang \etal 2005) and M31 (Mackey \etal 2010).

In 1980's there was a prediction that globular clusters are stripped off from galaxies in a galaxy cluster so that there should be a cluster-wide population of globular clusters that
are controlled by the gravitational potential of the galaxy cluster, rather than by that of any single galaxy (White 1987, Muzzio 1987, West \etal 1995).
These objects are called intracluster globular clusters (IGCs).
Then some studies came out, addressing the existence of IGCs in a few nearby galaxy clusters (e.g., Durrell \etal 2002, Tamura \etal 2006, Williams \etal 2007).
However, all these studies covered only a small fraction of a galaxy cluster prividing no information on the global distribution of IGCs in a galaxy cluster.

The Virgo cluster, the nearest galaxy cluster, is an ideal target to search for IGCs, but it spans a huge area in the sky, requiring very wide surveys.
Recently we created for the first time a surface number density map of globular clusters in the entire Virgo cluster using the SDSS data (Adelman-McCarthy \etal  2008), as shown in Fig. 1(a) (Lee \etal 2010a).
Fig. 1(b) and 1(c) display,respectively, the surface number density map of early-type galaxies derived using the Virgo galaxy catalog (Binggeli \etal 1985), and the contour map of X-ray emission (B\"orhinger \etal 1994) for comparison.
Fig. 1 shows two notable features. First, globular clusters are strongly concentrated around bright early-type galaxies, but their distribution is much larger than the size of galaxies.
Second, there is a diffuse large scale structure of IGCs, covering a significant fraction of the Virgo cluster. 
It is roughly similar to those of the early-type galaxies and X-ray emission, but showing some notable differences as well.

\begin{figure}
\resizebox{0.99\columnwidth}{!}{%
\includegraphics{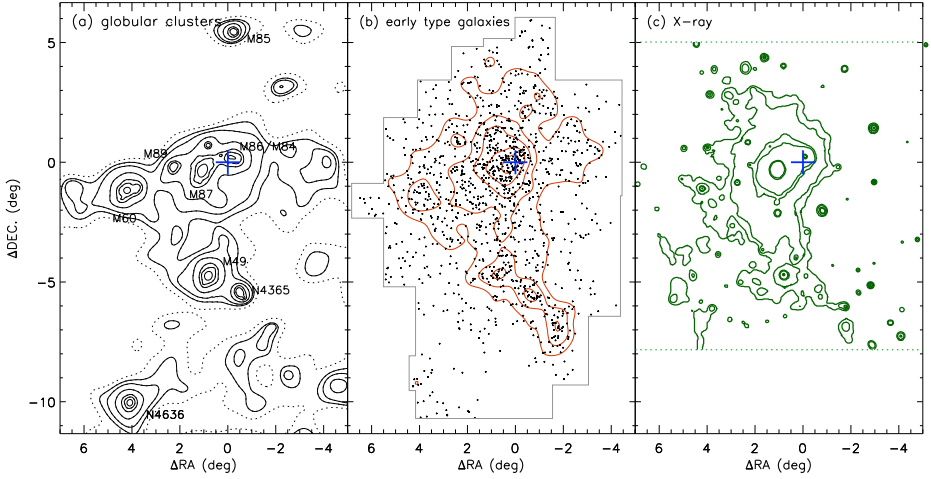}} 
\caption{The Virgo cluster. (a) The surface number density map of globular clusters. (b) The spatial distribution (dots) and surface number density contour map of early-type galaxies (89 E/S0s, 1142 dEs, and 39 dS0s). (c) The contour map of X-ray emission. The large cross represents the center of Virgo derived from the galaxy distribution.}
\label{fig1}       
\end{figure}

Fig. 2 displays a radial profile of the number density of globular clusters as well as that of dwarf galaxies in Virgo. All galaxies in Virgo were masked out except for M87 before deriving the number density profile. We adopted the center of M87 as the center for deriving the number density profile.
The radial profile of the globular clusters is dominated by the M87 globular clusters in the inner region ($R< 40'$), but it becomes flatter at $R\approx 40'$, showing that it is dominated by the IGCs.  Most of these IGCs are blue populations. The radial profile of these IGCs is similar to that of the dwarf galaxies that is also fitted well by the NFW profile.
These IGCs are probably dominated by the potential of the galaxy cluster or substructures in Virgo, rather than by that of a single galaxy. 
There are indeed wandering globular clusters in Virgo!

\begin{figure}
\resizebox{0.8\columnwidth}{!}{%
\includegraphics{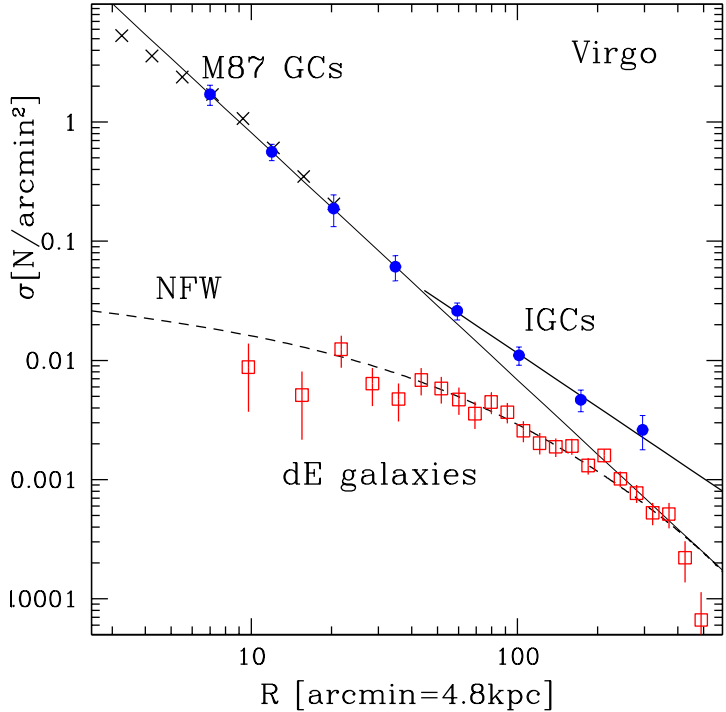}} 
\caption{The radial profiles of the number density of the globular clusters (filled circles) and dwarf galaxies (squares) in Virgo. Crosses represent the data for M87 globular clusters given by Harris (2009), and solid lines present the power-law fit to the globular cluster data
in the inner region ($R<40'$) and outer region ($R>40'$). The dashed line represents the NFW profile fit to the dwarf galaxy data. }
\label{fig2}       
\end{figure}

Fig. 3 displays images of some of these objects located around the Virgo center derived from the HST archive. Some resolved stars are seen around the outer part of these objects, showing that they are genuine globular clusters.
 
\begin{figure}
\resizebox{0.8\columnwidth}{!}{%
\includegraphics{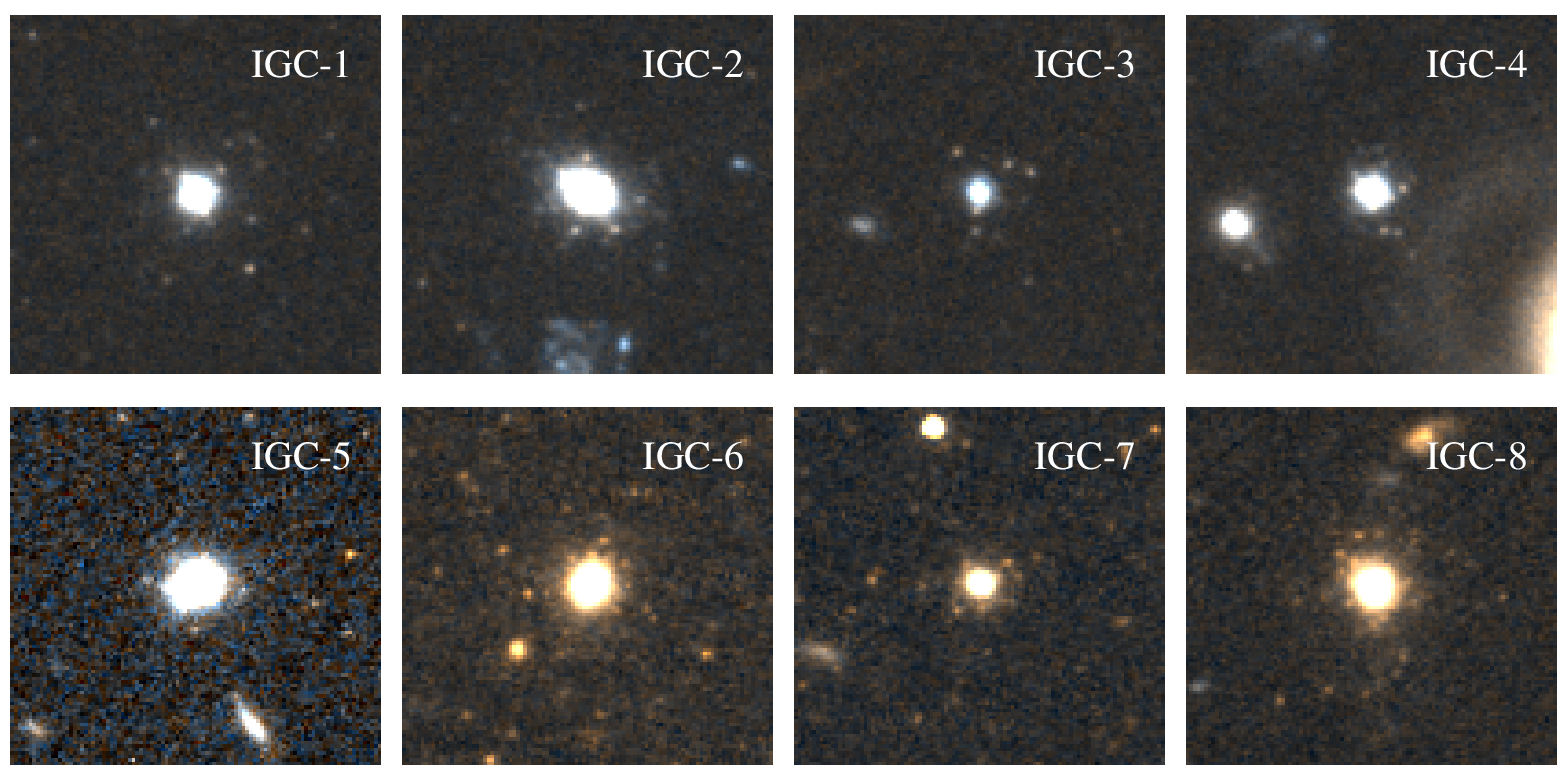}} 
\caption{HST images of some IGCs in the central region of the Virgo cluster. The objects in the upper row are previously known IGCs (Williams \etal 2007), and those in the lower row are our new findings 
(Lee \etal 2010c).}
\label{fig3}       
\end{figure}

\section{Implications}

Our discovery of the large scale structure of wandering globular clusters in Virgo has two implications on the dwarf systems in the universe:
(a) the origin of the globular clusters and (b) the first dwarf galaxies in the universe.

\subsection{The origin of the globular clusters}

Several models have been suggested to describe the formation of globular clusters  in the literature. 
These models can be broadly divided into five categories: 
 the first object model, the monolithic collapse model, the major merger model, the multiphase
  dissipational collapse model, and the dissipationless accretion model (see a summary in Lee 2003 and Lee \etal 2010b).
 
There are two recent findings useful for constraining
the globular cluster formation models. One is the discovery of the large scale structure of IGCs in Virgo as described above. The other is the kinematics of globular cluster systems in giant elliptical galaxies (gEs). Analysis of globular clusters in several gEs showed (a) that the kinematic properties of the globular cluster systems are diverse among gEs and
(b) that some kinematic parameters of the globular cluster systems show strong correlations with the global
parameters of their host galaxies (Lee \etal 2010b).

Considering these and other observational aspects of the globular clusters (Lee 2003, Brodie \& Strader 2006), 
Lee \etal (2010b) suggested a bibimbap (mixture) scenario for the origin of GCs in gEs as follows.
(1) Metal-poor globular clusters are formed mostly in low-mass dwarf galaxies very early, and preferentially in dwarf galaxies located in the high density environment like galaxy clusters. 
 These are the first generation of globular clusters in the universe. 
(2) Metal-rich globular clusters are formed together with stars in massive galaxies or dissipational merging galaxies later than metal-poor counterparts, but not much later. The chemical enrichment of the galaxies is rapid after the formation of metal-poor globular clusters.
(3) Massive galaxies grow becoming gEs via dissipationless or dissipational  merging of galaxies of various types and via accretion of many dwarf galaxies. New metal-rich globular clusters will be formed during dissipational merging, but the fraction of dissipational merging at this stage should be minor. 

In this scenario each gE has a different history of growing involved with diverse merging and accretion,
leading naturally to the diverse kinematics of the globular cluster systems in gEs. 
This scenario also explains various observational results: the bimodal color distribution of the globular clusters, the difference in spatial distribution between the blue globular clusters and red globular clusters, the correlation  in color between the red globular clusters and their host  galaxies, and the existence of wandering globular clusters.

\subsection{The first dwarf galaxies in the universe?}

A significant fraction of metal-poor globular clusters in gEs we see today are from dissipationless merging or accretion. During the accretion stage, many dwarf galaxies are completely disrupted. Some of the globular clusters that were in those galaxies are captured by massive galaxies and some of them are wandering far from any galaxies. Some of the dwarf galaxies are partially disrupted, leaving only their nucleus that will become exotic star clusters.
Therefore wandering globular clusters must be closely
related with the first galaxies in the universe.

\section{Future}

However, due to the shallowness of the SDSS data, we could use only bright globular clusters with $g<22.6$, which is one magnitude brighter than the turnover magnitude
of the globular cluster luminosity function (covering only 13\% of the total globular clusters).
Also we could not select directly individual globular clusters from the SDSS data, 
because we could derive only a statistical surface density map of the globular clusters. 
The number of IGCs identified in the HST images is too small.
Next step will be to select directly a large number of these globular clusters and to study their nature. 

C\^ot\'e \etal (2001) found that the velocity dispersion of the globular clusters in the outer area of M87 
increases as the galactocentric distance increases
and pointed out that the kinematics of the globular clusters in the outer region of M87 is affected significantly by the potential of the Virgo cluster rather than M87 itself. It is expected that the kinematics of IGCs found much farther from M87 will be dominated by
the cluster potential. Therefore these IGCs will be useful to study the dynamical history of the galaxy cluster and to trace the dark matter in Virgo.

\bigskip
MGL was supported by Mid-career Researcher Program 
through NRF grant funded by the MEST (No.2010-0013875).





\begin{thebibliography}{99}
\bibitem[2008]{ade08}Adelman-McCarthy, J.K. \etal 2008, ApJS, 175, 297
\bibitem[2006]{big85}Binggeli, B. \etal 1985, AJ, 90, 1681
\bibitem[2006]{bor94} B\"orhinger, H. \etal 1994, Nature, 368, 828 
\bibitem[2006]{bro06} Brodie, J.P. \& Strader, J. 2006, ARAA, 44, 193 
\bibitem[2001]{cot01} C\^ot\'e, P. \etal\ 2001, ApJ, 559, 828
\bibitem[2002]{dur02} Durrell, P.R. \etal 2002, ApJ, 570, 119
\bibitem[2009]{har09} Harris, W. 2009, ApJ, 703, 939
\bibitem[2005]{hwa05}Hwang, N. \etal 2005, Near-fields cosmology with dwarf elliptical galaxies, IAU Colloquium Proceedings 198, edited by Jerjen, H.; Binggeli, B. Cambridge: Cambridge University Press, 257
\bibitem[2003]{lee03} Lee, M.G. 2003, Jour. Korean Astron. Sco., 36, 189
\bibitem[2010]{lee10a} Lee, M.G. \etal 2010a, Science, 328, 334
\bibitem[2010]{lee10b} Lee, M.G. \etal 2010b, ApJ, 709, 1083
\bibitem[2010]{lee10c} Lee, M.G. \etal 2010c, in preparation
\bibitem[2010]{mac2010} Mackey, A.D. \etal 2010, ApJL, 717, L11
\bibitem[1987]{muz95} Muzzio, J.C. 1987, PASP, 99, 245 
\bibitem[2006]{tam06} Tamura, N. \etal  2006, MNRAS, 373, 601
\bibitem[1956]{van56} van den Bergh, S. 1956, PASP, 68, 449  
\bibitem[1995]{wes95} West, M. \etal 1995, ApJ, 453, L77
\bibitem[1987]{whi87} White, R.E. III 1987, MNRAS, 227, 185 
\bibitem[2002]{wil02} Williams, B.F. \etal 2002, ApJ, 654, 835


\end{thebibliography}
\end{document}